\documentclass[11pt,twoside]{article}

\usepackage{asp2014}

\aspSuppressVolSlug
\resetcounters

\begin{document}

\title{Insights from a 30-Year international Partnership on Astronomical Archives}

\author{David~R.~Rodriguez,$^1$ M.~Arevalo,$^2$ P.~Dowler,$^3$ J.~Espinosa,$^2$ B.~McLean,$^1$ and C.~Willot$^3$}
\affil{$^1$Space Telescope Science Institute, Baltimore, MD, USA; \email{drodriguez@stsci.edu}}
\affil{$^2$Starion Spain for ESA, Madrid, Spain}
\affil{$^3$CADC, National Research Council Canada, Victoria, BC, Canada}

\paperauthor{David~R.~Rodriguez}{drodriguezl@stsci.edu}{0000-0003-1286-5231}{Space Telescope Science Institute}{CSB}{Baltimore}{MD}{Postal Code}{USA}
\paperauthor{María~Arévalo}{Maria.Arevalo@ext.esa.int}{0000-0001-8102-0916}{Starion Spain for ESA}{SCI-MS}{Villanueva de la Cañada}{Madrid}{28692}{Spain}
\paperauthor{Patrick~Dowler}{patrick.dowler@nrc-cnrc.gc.ca}{0000-0001-7011-4589}{National Research Council}{CADC}{Victoria}{BC}{V9E 2E7}{Canada}
\paperauthor{Javier~Espinosa}{javier.espinosa@ext.esa.int}{0000-0002-9956-4433}{Starion Spain for ESA}{SCI-MS}{Villanueva de la Cañada}{Madrid}{28692}{Spain}
\paperauthor{Brian~McLean}{mclean@stsci.edu}{0000-0002-8058-643X}{Space Telescope Science Institute}{CSB}{Baltimore}{MD}{Postal Code}{USA}
\paperauthor{Chris~Willott}{Christopher.Willott@nrc-cnrc.gc.ca}{0000-0002-4201-7367}{National Research Council}{CADC}{Victoria}{BC}{V9E 2E7}{Canada}



\begin{abstract}
In an era where astronomical data is expanding at an unprecedented rate, the importance of data sharing and accessibility among astronomy archives cannot be overstated. 
Since the 1990s, an international partnership between the Space Telescope Science Institute (STScI), the European Space Astronomy Centre (ESAC), 
and the Canadian Astronomy Data Centre (CADC) has been focused on this endeavor, facilitating the exchange of data from the Hubble and James Webb Space Telescopes.

We will present how this collaboration has evolved over time, highlighting key milestones and innovations in decision-making, communication, and technology. 
Additionally, we will discuss some of the challenges we have encountered and the strategies we employed to overcome them, offering insights that could benefit future archive collaborations.
\end{abstract}

\section{Introduction}

When working towards supporting an archive for the Hubble Space Telescope, we had three main partners that worked to make this a reality.
These are the Space Telescope Science Institute (STScI), the European Space Astronomy Centre (ESAC), and the Canadian Astronomy Data Centre (CADC).

The Space Telescope Science Institute (STScI) is located at Baltimore, MD, USA and was established in 1981 and the Mikulski Archive for Space Telescopes (MAST) established in 1997 to serve IUE and HST data. MAST currently hosts data from over 20 missions, with a focus on UV/Optical/Near-IR observatories.

The Canadian Astronomy Data Centre (CADC) is located near Victoria, Canada and was established in 1986.
CADC serves active Canadian Space Agency (CSA) missions (including JWST), HST and many ground-based telescope archives.
CADC also runs the CANFAR Science Platform for collaborative science analysis of JWST and other missions.

The European Space Astronomy Centre (ESAC), located near Madrid, Spain, was established in 2002. Serves data from ESA's space science missions (including HST \& JWST) The archival partnership between NASA and ESA began in the early 1990s through the Space Telescope European Coordination Facility (ST-ECF), which operated at the European Southern Observatory (ESO) in Garching, Germany, until 2011. Following the closure of the ST-ECF, the European Hubble Archive was transferred to the ESAC Science Data Centre (ESDC).

\section{Collaboration History}

When we started working on sharing data, we focused on all public HST data. 
With the start of "On-the-Fly-Calibration", we switched sharing only raw data allowing each archive to run their own pipeline. 
We continued to collaborate at the instrument level to help design the pipeline, build products, and address any issues.

During the development of the Hubble Legacy Archive (HLA), responsibilities for processing data from various instruments and observation modes were distributed among different archives. STScI managed the processing of ACS and NICMOS imaging data; CADC handled WFPC2 data; while ST-EFC and ESA processed ACS and NICMOS grism data through 2010. 

The need for a more robust mechanism for data sharing became evident in the 2010s due to limitations  with the On-the-Fly Calibration approach. 
Because each archive had their own implementation of the pipeline, differences could arise depending on software versions, calibration reference files, or even hardware differences. 
As such, there was no single official best reduction of any HST product. 
In addition, each partner generated their own set of metadata to describe the data and provide search capabilities to their users and this provided an inconsistent description and experience.

This spurred the design of the HST Partner Consolidated Pipeline Project, which we describe in the following section.

\section{Consolidated Pipeline Project}

The Consolidated Pipeline Project was designed to handle the discrepancies of HST reduction across the three partners \citep{2019ASPC..523..425D}. 
The goal was to have a single archive run the pipeline and produce the highest level products. 
The other two archives would become data mirrors and harvest all the observation metadata and public data. 
This ensures that all three archives host the same copy of the products. 

In our setup today, STScI runs the pipelines to reduce both HST and JWST observations. 
CADC and ESAC harvest metadata and public data for these missions, roughly every hour. 
ESAC provides access to proprietary data through their interface via a redirect back to STScI.

\subsection{Tools}

To facilitate the sharing of metadata across archives we made use of the Common Archive Observation Model (CAOM; https://github.com/opencadc/CAOM/). 
CAOM was developed by CADC to handle metadata for astronomical observations across the electromagnetic spectrum. 
In CAOM, files are grouped as products of an Observation and elements, such as calibration levels, filter information, footprint formats, etc, are standardized. 
Because of this standardization, multiple missions can be searched with the same keywords empowering multi-mission archives. 
It is also extensible with supplementary information such as providing mission-specific instrument keywords, target keywords, and more. 
A working draft is available and CAOM is on track to become an IVOA standard. 

Another major tool is the the Metadata \& File harvesting software, which focuses on 
transferring the metadata (CAOM Observations in XML format) and the actual mission files themselves. 
These were originally developed by CADC, with contributions from ESAC, 
and are available as open source tools at https://github.com/opencadc.
These tools are installed at CADC and ESAC and allow us to list and get both metadata and files from a running CAOM service at STScI. 

\subsection{Communication}

We have regular check ins to continue to foster this partnership. 
We have a shared Slack channel for asynchronous communication and discussions. 
We also host bi-weekly stand-up meetings to check on status and go over any major issues. 
These meetings are hosted via Zoom at a time that works for all three partners. 
Finally, we have an annual Archive Coordination Meeting (ACM). 
The ACM typically lasts 2 days and we go over major points and plan out future work. 
For more long-term record keeping, STScI hosts an Outerspace website to store meeting notes and other documentation.

In addition to these communication channels, we also have automated tests that validate the content of each other's archives. 
These either send emails or slack messages for us to review.
This has been particularly useful at identifying discrepancies or delays in harvesting across the archives. 

\section{Lessons Learned}

Here we list the various lessons learned throughout this 30+ year partnership.

\textbf{Agree on standards to use}. 
We decided to use Common Archive Observation Model (CAOM) to handle metadata sharing. 
The metadata serialized as XML files for ease of transfer. 
We keep having discussions on best way to translate mission data into CAOM. 
And we make use of the Table Access Protocol (TAP) to check status across the archives.

\textbf{Communicate expectations}.
Early on we agreed to mirror only public data (eg, no proprietary/exclusive access data). 
We make sure to communicate timelines, data volume, ingest rates, and what we expect to see for them. 

\textbf{Share tools and resources}. 
CAOM was developed by CADC and integrated into STScI's pipeline.
The same harvester software is used by CADC and ESAC. 
STScI is also exploring how to use the harvester for internal testing.

\textbf{Have established check ins}.
We have bi-weekly stand-up meetings with the core team to ensure we are on track. 
We also have yearly meetings with a broader group to highlight status and work towards future goals.

\textbf{Have a robust testing infrastructure}.
We have a variety of Python and Java validation tools for CAOM content. 
Each archive checks metadata independently, but also do cross-archive checks. 
We've created test databases with sample HST, JWST content for checks prior to installing pipelines in production machines.
We're working towards producing dashboards for quick verification of results.

\textbf{Document the process}.
Given the long span of this collaboration, we know that systems and staff change over time. 
Capturing how things work can make it easier to make changes in the future and train new staff.

\textbf{Enable future collaborations}.
We continue to build trust and confidence on top of the collaboration to enable future projects.

\acknowledgements 

A large number of staff have participated in this project over the year. 
Here we acknowledge their contributions and list them.

At CADC, Patrick Dowler, Chris Willott, Daniel Durand, Severin Gaudet, JJ Kavelaars, David Schade, Ling Shao, and Hossen Teimoorinia.

At ESAC, María Arévalo, Javier Espinosa, Bruno Merín, Christophe Arviset, Deborah Baines, Anthony Marston, Christopher Evans, Paule Sonnentrucker, Antonella Nota, Raúl Gutiérrez, Javier Durán, Felix Stoehr (former ST-ECF) and Daniel Lennon.

At STScI, Brian McLean, David Rodriguez, David Wolfe, Matt Burger, Faith Abney, Rick White, Travis Berger, Josh Peek, Susan Mullally, Jonathan Hargis, and Karen Levay.

\bibliography{C502}  

\begin{thebibliography}{}
\expandafter\ifx\csname natexlab\endcsname\relax\def\natexlab#1{#1}\fi
\expandafter\ifx\csname url\endcsname\relax
  \def\url#1{\texttt{#1}}\fi
\expandafter\ifx\csname urlprefix\endcsname\relax\def\urlprefix{URL }\fi
\providecommand{\eprint}[2][]{\url{#2}}

\bibitem[{{Dowler} et~al.(2019){Dowler}, {Arevalo}, {Damian}, {Duran},
  {Durand}, {Gaudet}, {Hargis}, {Major}, {McLean}, {Oberdorf}, \&
  {Rodriguez}}]{2019ASPC..523..425D}
{Dowler}, P., {Arevalo}, M., {Damian}, A., {Duran}, J., {Durand}, D., {Gaudet},
  S., {Hargis}, J., {Major}, B., {McLean}, B., {Oberdorf}, O., \& {Rodriguez},
  D.~R. 2019, in Astronomical Data Analysis Software and Systems XXVII, edited
  by P.~J. {Teuben}, M.~W. {Pound}, B.~A. {Thomas}, \& E.~M. {Warner}, vol. 523
  of Astronomical Society of the Pacific Conference Series, 425

\end{thebibliography}


\end{document}